\documentclass{ckm}                 

\def\F{{\cal F}}
\newcommand{\xba}{\bar}
\newcommand{\ETA}{\zeta}

\confname{Workshop on the CKM Unitarity Triangle, IPPP Durham, April
  2003}

\title{Measurement of $\gamma$ at B Factories Using Inclusive D Decays}

\author{
\underline
{ D.~Atwood\addressmark{a}} and A.~Soni\addressmark{b}}
\address[a]{Dept. of Physics, 
Iowa State University, Ames, Iowa, 50011 USA}
\address[b]{Theory Group, Brookhaven National 
Laboratory, Upton, NY 11973 USA}

\begin{document}

\begin{abstract}
We discuss the determination of the CKM phase $\gamma$ through the decay
$B^-\to K^- D^0$ and related processes. In particular, we consider the use
of this methods when the $D^0$ subsequently decays to an inclusive
state. We emphasize that strong phase information obtained at a
$\psi(3770)$ charm factory can provide additional information which will
be helpful in determining $\gamma$.

\end{abstract}

\maketitle

\section{Introduction}

One of the most important missions of the B factories is to carry out
precision tests of the Standard Model. Among the key predictions is that
the Cabibbo Kobayashi-Maskawa (CKM) matrix is unitary~\cite{ckm}. This
fact is most
elegantly expressed by the closing of the unitarity triangle. 

CP violation in $B$ decays allow the separate measurement of the three
angles of the unitarity triangle.  The most successful such measurement to
date~\cite{belle_beta,babar_beta} 
is the determination of $\sin 2\beta$. A crucial point to note is
that this result is subject to no significant theoretical error so
the precision will improve with more data.

In this talk I will consider the determination of $\gamma$ which has the
same property, that there is little theoretical error. The basic
idea~\cite{glw,ads} is to interfere the quark level process $b\to c \bar u
s$ with $b\to u \bar c s$. Of course this interference can only take place
if the final states are the same which can be achieved in the decay
$B^-\to K^- D^0$ interfering with $B^-\to K^- \bar D^0$ through decays of
$D^0$, $\bar D^0$ to a common final state.

Two classes of final states are of particular interest. First of all,
there are states which are CP eigenstates~\cite{glw} (CPES) such as $K_S
\pi^0$. Second of all, there are $D^0$ decay~\cite{ads} which are Doubly
Cabibbo suppressed (DCS) such as $K^+\pi^-$.
In addition, recent work~\cite{scs} has also considered in this context 
final states which are
singly Cabibbo suppressed but not CP eigenstates such as $K^*\bar K$.
In this talk I will focus on CPES and DCS final states.

For a given final state $f$, let us denote by $d$ the combined branching
ratio $B^-\to K^- [D^0\to f]$ and $\bar d$ the combined branching ratio
$B^+\to K^+ [D^0\to \bar f]$ (here $D^0$ represents a generic mixture of
$|D^0\rangle$ and $|\bar D^0\rangle$).
Thus,

\begin{eqnarray}
d(f)&=& ac_f + b\bar c_f +2\sqrt{ac_fb\bar
c_f}\cos(\zeta_B+\zeta_f+\gamma)
\nonumber\\
\bar
d(f)&=& ac_f + b\bar c_f +2\sqrt{ac_fb\bar
c_f}\cos(\zeta_B+\zeta_f-\gamma)
\label{exclusive_d}
\end{eqnarray}

\nonumber Where $c_f$ is the branching ratio of $D^0\to f$, $\bar c_f$ is
the branching ratio of $\bar D^0\to f$, $a$ is the branching ratio of
$B^-\to K^- D^0$ and $b$ is the branching ratio of $B^-\to K^- \bar D^0$.  
$\zeta_B$ is the strong phase difference between $B^-\to K^- D^0$ while
$\zeta_f$ is the strong phase difference between $D^0\to f$ and $\bar
D^0\to f$. Here we assume that $D\bar D$ mixing is small~\cite{ads,ss1}

This formalism, however, applies only to cases where the final state of
the $D^0$ decay is an exclusive state controlled by a single amplitude.
This is the situation if the $D^0$ decays to a two body final state with a
single helicity amplitude (e.g. $K^+\pi^-$) but will not be true for
decays to 3 or more final state particles. In such multi-body decay each
point in phase space behaves like an exclusive state but the integrated
rate over phase space we will regard as an inclusive state. Likewise we
can regard the combination of states with different particle content as
inclusive states.

In the following we will develop the formalism for such inclusive states
analogous to Eqn.~(\ref{exclusive_d}). This will allow for a strategy to
improve the determination of $\gamma$ with the addition of information
from a $\psi(3770)$ charm factory such as CLEO-c. The improvement results
primarily from the improved statistics which are obtained from the use of
inclusive states where the charm factory provides key information needed
to interpret that data.

\section{Formalism for Inclusive States}

Let us consider $B^-\to K^- [D^0\to F]$. Each of the states $f_i$ in $F$ 
satisfies Eqn.~(\ref{exclusive_d}). Summing over $i$, 
we obtain~\cite{deltapaper,ads_last,scs2}:

\begin{eqnarray}
d(F)&=& ac_F + b\bar c_F +2R_F\sqrt{ac_Fb\bar
c_F}\cos(\zeta_B+\zeta_F+\gamma)
\nonumber\\
\bar
d(F)&=& ac_F + b\bar c_F +2R_F\sqrt{ac_Fb\bar
c_F}\cos(\zeta_B+\zeta_F-\gamma)
\nonumber\\
\label{inclusive_d}
\end{eqnarray}

\noindent
where $R_F$ and $\zeta_F$ are defined by:

\begin{eqnarray}
R_F e^{i\zeta_F} 
={
\sum_i 
\sqrt{ c_{f_i}  \bar c_{f_i}} e^{\zeta_i}
\over
\sqrt{ c_F  \bar c_F}
}
\end{eqnarray}

\noindent
We can regard $\zeta_F$ as the average strong phase of the inclusive set
of states, $F$, and $R_F$ as the coherence of this set of states.

\begin{figure}
\hbox to\hsize{\hss 
\includegraphics[width=\hsize]
{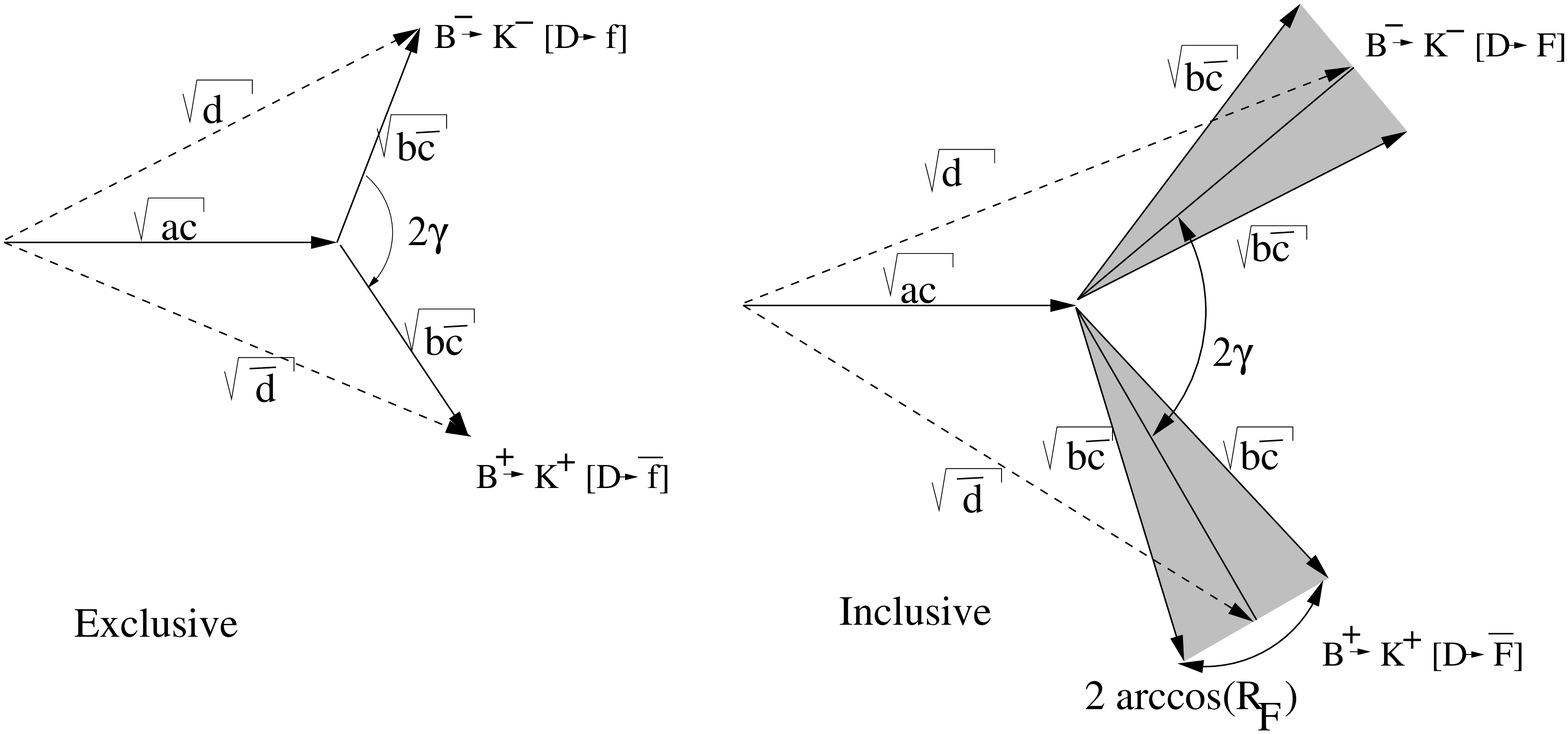} \hss}
\caption{
The geometric representation of Eqn.~(\ref{inclusive_d}) and
Eqn.~(\ref{exclusive_d}).
} 
\label{atwood_fig1}
\end{figure}

Comparing Eqn.~(\ref{inclusive_d}) with Eqn.~(\ref{exclusive_d}) we see
that the cross section for inclusive states requires an additional
parameter, $R_F$, as compared to the case of an exclusive state. This can
be understood in terms of the diagrams in Fig.~(\ref{atwood_fig1}) where
we show the geometric relation between the amplitudes of the two channels
as vectors in the complex plane. In the exclusive case we take
the phase convention that the amplitude via the $b\to c$ channel is real
while the $b\to u$ channel has a combination of strong and weak phases.
The vector representing the $b\to u$ channel thus swings through $2\gamma$
when moving from the $B^-$ to the corresponding $B^+$ decay.  In the
inclusive case, the second leg of each triangle opens up into a cone of
angle $2\arccos R_F$ because of the incoherence of the strong
phases.
From this it can be clearly seen that if $R_F=1$ then the situation
reduces to the exclusive case while if $R_F=0$ the there is no CP
violation.

As was pointed out by~\cite{glw}, that if you know $d$, $\bar d$, $ac$ and
$b\bar c$ you can reconstruct the diagram (up to discrete ambiguities)  
and thus determine $\gamma$. In the exclusive case, however, you would
also have to know $R_F$.

Even with exclusive states, this approach may not be practical because the
determination of the small branching ratio $B^-\to K^- \bar D^0$ is
difficult~\cite{ads}.  This can be circumvented if two distinct exclusive
states are used since then one does have enough information to solve for
all the strong phases and $\gamma$.  This fails in the case of inclusive
states due to the extra parameter, $R_F$.

At a $\psi(3770)$ charm factory~\cite{charmfactory} it is possible to
determine
$\zeta_f$ for an exclusive state~\cite{sofferun} and $\zeta$ and $R_F$
for an inclusive final state~\cite{deltapaper,ads_last}. This is because
the decay
$\psi(3770)\to D^0\bar D^0$ leaves the final state $D^0$'s in an
anti-symmetric flavor state therefore if one of the $D$-mesons decays to
an
inclusive state $F$ and the other to state $G$, the correlated
rate~\cite{ads_last}
(again, assuming $D$ mixing is small~\cite{ss1}) is:

\begin{eqnarray}
\Gamma(FG)
&=&
\Gamma_0   
\bigg[
A_F^2\xba A_G^2
+
\xba A_F^2  A_G^2
\nonumber\\
&&-2 R_F
R_G
A_F
\xba A_F  
A_G
\xba A_G
\cos(\ETA_F - \ETA_G)
\bigg]
\label{FG} 
\end{eqnarray}

\noindent
where $\Gamma_0=\Gamma(\psi(3770)\to D^0\xba D^0$. If $G$ is a 
CPES with CP eigenvalue $\lambda_G=\pm 1$, then this reduces to:

\begin{eqnarray}
\Gamma(FG)
&=&   
\Gamma_0A_G^2
\bigg[
A_F^2
+
\xba A_F^2
-2 \lambda_{G} R_F
A_F
\xba A_F
\cos(\ETA_F)
\bigg]  
\nonumber\\
\label{FCP}
\end{eqnarray}

%
%
%
%

\noindent
and if $F=G$ then

\begin{eqnarray}
\Gamma(FF)   
&=&
\Gamma_0
A_F^2\xba A_F^2(1-R_F^2)
\label{FF}
\end{eqnarray}

Note in this latter case that if $R_F=1$ then $\Gamma(FF)=0$ by Bose
symmetry.

Thus, let us suppose that $n$ distinct inclusive states as well as CP
eigenstates are considered and all correlations are measured at a
$\psi(3770)$ charm factory. 
The number of unknown  
parameters is $2n$ because there is $R_F$ and $\eta_F$ for each of the
inclusive states. The number of correlations which can be measured is
$n^2+2n$ as follows:

\begin{enumerate}
\item Each state can be correlated with any other state including itself
giving $n(n+1)/2$ observables.
\item  Each state can be correlated with the charge conjugate of any other
state including itself
giving $n(n+1)/2$ observables.
\item
Each state can be correlated with a CPES 
giving $n$ observables.
\end{enumerate}

There are therefore $n^2$ more observables than parameters. The same is
also true if some of states are exclusive since an exclusive state has
only one parameter ($\eta_f$) but the branching ratio $\psi(3770)\to ff$
is also automatically 0 by Bose symmetry so there is also one less
observable.

\section{Illustrative Example}

Let us now consider some examples to illustrate how various
combinations of final states can be taken together to determine $\gamma$.

First of all, let us consider two exclusive final states $f_1=K^+\pi^-$
and $f_2=K^{*+}\pi^-$ as well as CP=-1 eigenstates.  The branching ratio
$c(f_1)=1.48\times 10^{-4}$~\cite{pdb} while $\bar c(f_1)=3.80\times
10^{-2}$~\cite{pdb};  $c(f_2)=1.7\times 10^{-4}$ while $\bar
c(f_1)=6.0\times 10^{-2}$~\cite{pdb}. Note $c(f_2)$ is not yet measured so
we estimate it by multiplying $\bar c(f_2)$ by $\tan^4\theta_C$.  For
illustrative purposes, we will take the ad-hoc values
$\zeta_{f_1}=120^\circ$ and $\zeta_{f_2}=120^\circ$. We will take the
branching ratio of $CPES-$ states to be $\sim 5\%$, roughly the rate of
$D^0$ 2 body decays to $K_S$ and a pseudo-scalar or vector.

\begin{figure}
\hbox to\hsize{\hss 
\includegraphics[width=\hsize]
{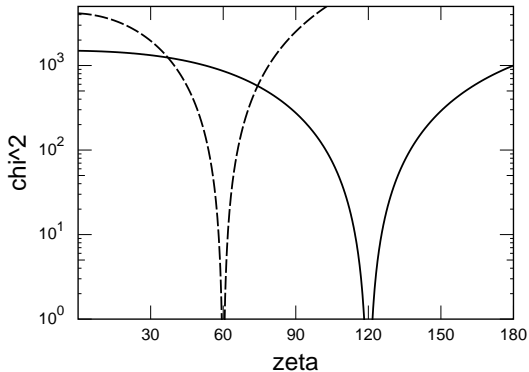} \hss}
\caption{
$\chi^2$ is plotted as a function of $\ETA$
using correlations at a $\psi(3770)$ factory using
correlations between the exclusive modes considered in
$f_1$ and $f_2$. The solid line is for the mode $f_1=K^+\pi^-$ while
the dashed line is for the mode $f_2=K^{*+}\pi^-$.
} 
\label{atwood_fig2}
\end{figure}

In Fig.~(\ref{atwood_fig2}) we show the minimum value of $\chi^2$ at each
possible value of $\zeta$ for each of the two exclusives states were we
use the 6 possible correlations as discussed above (i.e. $f_1\times \bar
f_1$, $f_2\times \bar f_2$, $f_1\times f_2$, $f_1\times\bar f_2$,
$f_1\times CPES$, $f_2\times CPES$). To calculate $\chi^2$ we take $\hat
N_D=(acceptance)\times (number~of~D^0\bar D^0)= 10^7$. Probably $\hat
N_D\sim 100-300$ would roughly correspond to reasonable $3\sigma$ errors
form CLEO-c.

Let us now see how well this can be used to determine $\gamma$. For the
purpose of illustration, we will assume that $\zeta_B=-50^\circ$ and
$\gamma=60^\circ$, which is consistent with the current CKM
fits~\cite{ckmfit}. 

\begin{figure}
\hbox to\hsize{\hss 
\includegraphics[width=\hsize]
{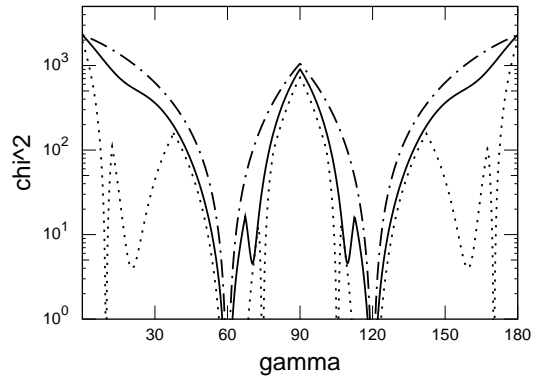} \hss}
\caption{
$\chi^2$ is plotted as a function of $\gamma$ for input generated assuming
$\gamma=60^\circ$ and $\zeta_B=-50^\circ$ using various combinations of
exclusive states with $\hat N_B=10^9$. The dotted curve uses only data
from two $D$ decay modes: CPES- and $K^{*+}\pi^-$; the solid curve
includes data from three modes: CPES-, $K^{*+}\pi^-$ and $K^-\pi^+$. The
dot-dashed curve uses the same modes but correlation data from a
$\psi(3770)$ factory with $\hat N_{D}=10^7$ is also included. 
} 
\label{atwood_fig3}
\end{figure}

In Fig.~(\ref{atwood_fig3}) we show the minimum $\chi^2$ as a function of
$\gamma$ for a number of combinations of the data sets assuming 
$\hat N_B=(number~of~B's)\times (Acceptance)=10^9$. Notice that the 
curves are symmetric with respect to $\gamma\to\pi-\gamma$ as well as
$\gamma\to\pi+\gamma$ therefore there is ultimately a 4-fold ambiguity in
the determination of $\gamma$ through this method.

The dotted curve shows the results where we just use $d$ and $\bar d$ for
both $f_2=K^{*+}\pi^-$ and $CPES-$. Notice that the curve is pulled to 0
at a number of points due to the discrete ambiguities of the solution in
this case because the four observables $d(CPES-)$, $\bar d(CPES-)$,
$d(f_2)$, $\bar d(f_2)$ are determined by the four unknowns $b$, $\gamma$,
$\eta_B$ and $\eta_{f_2}$. If we add another mode, $f_1=K^-\pi^+$ as is
shown by the solid curve, the situation is improved since this adds two
more observables, $d(f_1)$ and $\bar d(f_1)$ but only one more unknown,
$\eta_{f_1}$. Clearly the ambiguities are largely lifted in fact. Finally,
if we fit the B factory data jointly with the $\psi(3770)$ data discussed
above we obtain the dash dotted curve where the parameters $\zeta_{f_1}$
and $\zeta_{f_2}$ are, in effect, determined from independently. Looking
at the $3\sigma$ errors of $\gamma$ estimated by taking the intersection
for the curve with $\chi^2=9$, we see that for the case of $f_1$ and
$CPES-$, the error is $10^\circ$ (although there is additional confusion
due to the multiple minima of the curve); for the case if $f_1$, $f_2$ and
$CPES-$ the error is $9^\circ$ and for $f_1$, $f_2$ and $CPES-$ with charm
factory data the error is $3.4^\circ$. In the last case the curve also
smoothly funnels down to the correct value of $\gamma$ indicating that the
convergence will scale well as data is improved.

\begin{figure}
\hbox to\hsize{\hss 
\includegraphics[width=\hsize]
{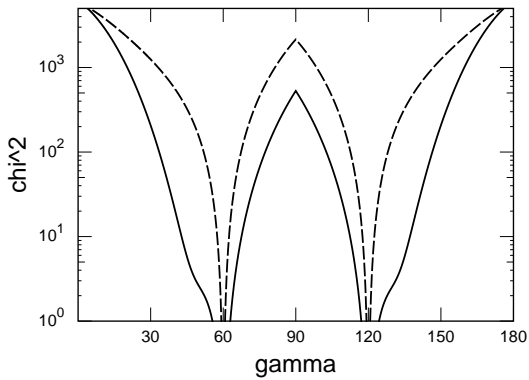} \hss}
\caption{
$\chi^2$ is plotted as a function of $\gamma$ for input generated assuming
$\gamma=60^\circ$ and $\zeta_B=-50^\circ$ for various combinations of
inclusive states with $\hat N_B=10^9$ together with $\psi(3770)$ factory
data. The solid curve uses only $\F$ and CPES-. The dashed line uses
information from $\F_{1-3}$ together with CPES-.
} 
\label{atwood_fig4}
\end{figure}

Let us turn our attention now to the an inclusive case where we consider
decays of the form $\F=\{ K^++X \}$. We will model this set of states
using the model discussed in~\cite{ads_last}. We shall also consider
breaking down this set according to the energy of the $K^+$ where $
\F_1$ has $E_K\leq 1.7GeV$, $\F_2$ has $1.7GeV\leq E_K\leq 4.7GeV$ and
$\F_3$ has $4.7GeV<E_K$. In this model, $\zeta_\F=-11^\circ$,
$\zeta_{\F_1}=-34^\circ$, $\zeta_{\F_2}=-86^\circ$ and
$\zeta_{\F_3}=-30^\circ$, while $R_\F=.51$, $R_{\F_1}=.74$, $R_{\F_2}=.29$
and $R_{\F_3}=.91$.

If we fit $\gamma$ using $d$ and $\bar d$ for $\F$ and $CPES-$ together
with charm factory data determining $R_\F$ and $\zeta_F$ we obtain the
solid curve in Fig.~(\ref{atwood_fig4}). The corresponding $3\sigma$ error
in $\gamma$ is $12^\circ$. If, however, we take the exact same data 
except we break $\F$ down into $\F_{1-3}$ we obtain the dashed curve in
Fig.~(\ref{atwood_fig4}) with the corresponding error of $2.3^\circ$. The
improvement is due to the fact that in breaking down the data we have
increased the number of constraints compared to the number of unknown
parameters. Of course the statistical errors on each of the $d$ and $\bar
d$'s in the latter case will be larger. The improvement of the latter case
with respect to the exclusive states is largely due to the fact that 
we have larger statistics overall. 
Clearly, as discussed in~\cite{ads_last,scs2}, it is important to consider
different strategies for breaking down inclusive data in order to obtain
the best determination of $\gamma$.

\section{Conclusion}
In conclusion, I have shown how the strong phase properties of an
inclusive decay of $D^0$ may be represented by two parameters, $R_F$ and
$\zeta_F$. Correlated decays in a $\psi(3770)$ charm factory provide a
means of obtaining these parameters. This information may be put to good
use in obtaining the CKM phase $\gamma$ via $B^-\to K^- D^0$ so that a
$\psi(3770)$ charm factory will proved invaluable information to interpret
this data from $B$ factories.

\end{document}